\documentclass[preprintnumbers,showkeys,superscriptaddress,showpacs,
nofootinbib,byrevtex,fleqn]{revtex4}   
\usepackage{amsmath,amsfonts,amssymb,amscd,amsxtra,amsthm}
\usepackage{graphicx}
\usepackage{bm}
\usepackage{epstopdf}
\begin{document}
\preprint{INHA-NTG-03/2011}
\title{$K^*\Lambda(1116)$ photoproduction and nucleon resonances}
\author{Sang-Ho Kim}
\affiliation{Department of Physics, Inha University, Incheon 402-751, Korea}  
\author{Seung-il Nam}
\affiliation{Research Institute of Basic Sciences, Korea Aerospace
University, Goyang 412-791, Korea}  
\author{Yongseok Oh}
\affiliation{School of Physics and Energy Sciences, Kyungpook National University, Daegu 702-701, Korea}  
\author{Hyun-Chul Kim}
\affiliation{Department of Physics, Inha University, Incheon 402-751, Korea}  
\begin{abstract}
In this presentation, we report our recent studies on the
$K^*\Lambda(1116)$ photoproduction off the proton target, using the
tree-level Born approximation, via the effective Lagrangian
approach. In addition, we include the nine (three- or four-star
confirmed) nucleon resonances below the threshold
$\sqrt{s}_\mathrm{th}\approx2008$ MeV, to interpret the discrepancy
between the experiment and previous theoretical studies, in the
vicinity of the threshold region. From the numerical studies, we
observe that the $S_{11}(1535)$ and $S_{11}(1650)$ play an important
role for the cross-section enhancement near the
$\sqrt{s}_\mathrm{th}$. It also turns out that, in order to reproduce
the data, we have the vector coupling constants
$g_{K^*S_{11}(1535)\Lambda}=(7.0\sim9.0)$ and
$g_{K^*S_{11}(1650)\Lambda}=(5.0\sim6.0)$.  
\end{abstract}

\maketitle
\section{Introdcution}
The strangeness meson photoproduction off the nucleon target is one of
the most well-studied experimental and theoretical subjects to reveal
the hadron production mechanisms and its internal structures, in terms
of the strange degrees of freedom, breaking the flavor SU(3) symmetry
explicitly. Together with the recent high-energy photon beam
developments in the experimental facilities, such as LPES2 at SPring-8
and CLAS12 at Jefferson laboratory~\cite{EXP}, higher-mass strange
meson-baryon photoproducitons must be an important subject to be
addressed theoretically for future studies on those reaction
processes. In the previous works~\cite{Oh:2006in,Oh:2006hm}, the
$K^*\Lambda(1116)$ photoproduction was investigated, the Born
approximation being used with the Regge contributions. In comparison
with the preliminary experimental data~\cite{Guo:2006kt}, the theory
reproduced the data qualitatively well, but the theoretical
cross-section strength was underestimated in the vicinity of the 
$\sqrt{s}_\mathrm{th}$. In the present talk, we want to report our
recent study to explain this discrepancy observed in the previous
work. Based on the theoretical framework as employed in
Ref.~\cite{Oh:2006hm}, we include the nucleon resonances in the
$s$-channel baryon-pole contribution. As for the nucleon resonances
$N^*$, we take into account nine of them, i.e. $P_{11}(1440,1/2^+)$,
$D_{13}(1520,1/2^-)$,  $S_{11}(1535,1/2^-)$,  $S_{11}(1650,1/2^-)$,
$D_{13}(1700,3/2^-)$,  $P_{11}(1710,1/2^+)$,  $P_{13}(1720,3/2^+)$,
$P_{11}(1440,1/2^+)$,  $D_{15}(1675,5/2^-)$,  and
$F_{15}(1680,5/2^+)$, in a full relativistic manner.  
\section{Theoretical framework and numerical results}
\begin{figure}[t]
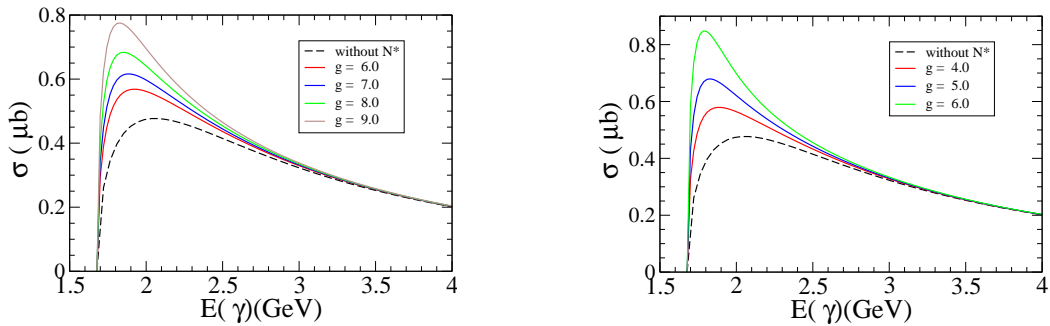

\begin{tabular}{cc}
\includegraphics[width=6cm]{fff1.eps}
\,\,\,\,\,\,\,\,\,\,\,\,\,\,\,\,\,\,\,\,\,\,\,\,\,\,\,\,\,\,
\includegraphics[width=6cm]{fff2.eps}
\end{tabular}
\caption{Total cross section for the $\gamma p \to
  K^{*+}\Lambda(1116)$ reaction process. We show the numerical results
  for the cases including the $S_{11}(1535)$ (left) and $S_{11}(1650)$
  (right), varying the coupling constants $g_{K^*N^*\Lambda}$.}        
\label{FIG1}
\end{figure}

We start with explaining the theoretical framework briefly and 
represent the important numerical results in our study.  We note that
the nucleon resonances are carefully taken into account in a
full-relativistic manner, in addition to the Born diagrams, $K$ and
$\kappa(800)$ meson-exchanges in the $t$ channel, and $\Sigma(1192)$
and $\Sigma^*(1385)$ hyperon exchanges in the $u$ channel, which were
already employed in Ref.~\cite{Oh:2006hm}. All the effective
interaction vertices for the nucleon resonances are taken from
Ref.~\cite{Oh:2007jd}. For instance, the invariant amplitudes for the
spin-$1/2$ and spin-$3/2$ resonance contributions in the $s$ channel
can be written as follows: 
\begin{eqnarray}
\label{eq:AMP}
\mathcal {M}_{N^*}\left(
  \frac{1}{2}^\pm\right)&=&\pm\frac{g_{{K^*}N^*\Lambda}}{s-M_{N^*}^2}
{\varepsilon}_{\nu}^*(k_2){\bar u}_{\Lambda}(p_2) 
\left[\gamma^{\nu}-{\frac{i\kappa_{K^{*}N^*\Lambda}}{2M_N}}
{\sigma^{\nu\beta}}k_{2\beta} \right]
\cr     
&&\frac{ie_Q\mu_{N^*}}{2M_N}\Gamma^{\mp}(\not{k_1}+\not{p_1}+M_{N^*})
\Gamma^{\mp}\sigma^{\mu{i}}k_{1i}u_N(p_1){\varepsilon}_{\mu}(k_1), \cr
\mathcal {M}_{N^*}\left(\frac{3}{2}^\pm \right)&=&
\frac{g_{{K^*}N^*\Lambda}}{s-M_{N^*}^2}
{\varepsilon}_{\nu}^*(k_2){\bar
  u}_{\Lambda}(p_2)({k_2^\beta}g^{{\nu}i}-k_2^ig^{\nu\beta}) 
\frac{e_Q}{2M_{K^*}}\Gamma_i^{\pm}\Delta_{\beta\alpha}         
\cr                                 
&&\left[\frac{\mu_{N^*}}{2M_N}\gamma_j\,\mp\,\frac{\bar \mu_{N^*}}
{4M_N^2}p_{1j} \right]\Gamma^{\pm}({k_1^\alpha}g^{{\mu}j}-{k_1^j}
g^{\mu\alpha})u_N(p_1){\varepsilon}_{\mu}(k_1),
\end{eqnarray}
where $(k_1,p_1,k_2,p_2)$ stand for the $(\gamma,p,K^*,\Lambda)$
momenta and $\mu_{N^*}$ for the helicity amplitude. The
$g_{K^*N^*\Lambda}$ and $\kappa_{K^*N^*\Lambda}$ denote the strong
vector and tensor coupling strengths, respectively. The polarization
vectors for the photon and $K^*$ are assigned as
$\varepsilon_\mu(k_1)$ and $\varepsilon_\mu(k_2)$. The $\Gamma$
controls the parity of the relevant resonances in the following way: 
\begin{equation}
\label{eq:}
 \Gamma_\mu^{\pm} = \left(
\begin{array}{c}
\gamma_\mu\gamma_5 
\\
\gamma_\mu
\end{array} \right),\,\,\,\,
\Gamma^{\pm} = 
\left(
\begin{array}{c} 
\gamma_5 \\ 
\bold 1_{4\times4}
\end{array} \right).
\end{equation}
Relevant parameters for the resonances are estimated using the
experimental and theoretical
information~\cite{Nakamura:2010zzi,Capstick:1998uh,Capstick:2000qj}. In
Figure~\ref{FIG1}, we show the total cross section for the present
reaction process, i.e. $\gamma p \to K^{*+}\Lambda(1116)$. The
numerical results are drawn separately for the cases including the
$S_{11}(1535)$ (left) and $S_{11}(1650)$ (right), varying the coupling
constants $g_{K^*N^*\Lambda}$. As shown in the figure, the
cross-section enhancement is observed in the vicinity of the threshold
region, if we choose the strong coupling strengths as
$g_{K^*N^*\Lambda}\approx(4.0\sim9.0)$. We verified that other nucleon
resonances are not so effective to interpret the discrepancy.   

\section{Summary and outlook}
In the present work, we have studied the $K^*\Lambda$ photoproduction
theoretically, employing the tree-level Born approximation and
nucleon-resonance contributions below the
$\sqrt{s}_\mathrm{th}$. Among the nucleon resonances, the
$S_{11}(1535)$ and $S_{11}(1650)$ play a dominant role to reproduce
the experimental data. It also turns out that other $N^*$
contributions are not so effective to improve the theoretical
results. We note that the nucleon resonances beyond the
$\sqrt{s}_\mathrm{th}$ may contribute to the threshold enhancement,
especially due to the $D_{13}(2080)$, since the $\sqrt{s}_\mathrm{th}$
for the present reaction process is about $2008$ MeV. Related works
are under progress and appear elsewhere.  
\section*{Acknowledgments}
The authors are  grateful to A.~Hosaka for fruitful discussions. The
present work is supported by Basic Science Research Program through
the National Research Foundation of Korea (NRF) funded by the Ministry
of Education, Science and Technology (grant number: 2009-0089525). The 
work of S.i.N. was supported by the grant 
NRF-2010-0013279 from National Research Foundation (NRF) of Korea.  

\end{document}